\documentclass[final,5p,times,twocolumn]{elsarticle}

\usepackage[table]{xcolor} 
\usepackage{amssymb}
\usepackage{amsmath}
\usepackage{multirow}
\usepackage{booktabs}
\usepackage{bm}
\usepackage{braket}
\usepackage{graphicx}
\usepackage[normalem]{ulem}
\biboptions{sort&compress}

\usepackage[pdfstartview=FitH, colorlinks,
 linkcolor={red!60!black!},
 anchorcolor={green!80!black!},
 urlcolor={blue!80!black!},
 citecolor={pink!90!black!}]{hyperref}

\journal{Physics Letters B}

\begin{document}

\begin{frontmatter}
\title{Probing exotic multi-proton emitters: A Gamow shell model study of proton-rich fluorine and neon isotopes beyond the drip line}
\author[ad1,ad2]{N. Chen}
\author[ad1,ad2,ad3]{J. G. Li \corref{correspondence}}
\author[ad1,ad2]{M. R. Xie}
\author[ad5,ad1]{P. Y. Wang}
\author[ad1,ad4]{K. H. Li}
\author[ad1,ad2]{Q. Yuan}
\author[ad1,ad2]{N. Michel}

\address[ad1]{State Key Laboratory of Heavy Ion Science and Technology, Institute of Modern Physics, Chinese
Academy of Sciences, Lanzhou 730000, China}

\address[ad2]{School of Nuclear Science and Technology, University of Chinese Academy of Sciences, Beijing 100049, China}

\address[ad3]{Southern Center for Nuclear-Science Theory (SCNT), Institute of Modern Physics, Chinese Academy of Sciences, Huizhou 516000, China}

\address[ad4]{College of Physics, Henan Normal University, Xinxiang 453007, China}

\address[ad5]{Advanced Energy Science and Technology Guangdong Laboratory, 516007, Huizhou, China}

\cortext[correspondence]{Corresponding author: jianguo\_li@impcas.ac.cn (J. G. Li)}

\begin{abstract}
We investigate proton-rich systems beyond the proton drip line, focusing on the notably poorly known $^{13}$F and $^{15}$Ne and the yet unobserved $^{14}$Ne, whose structure properties remain weakly constrained. 
Using the Gamow shell model (GSM), which consistently incorporates both inter-nucleon correlations and couplings to the particle continuum, we study oxygen, fluorine, and neon isotopes with mass $A=12-16$. 
Taking $^{8}$C as an inert core, the GSM Hamiltonian based on an effective field theory nucleon-nucleon interaction is optimized for this proton-rich region. The constructed Hamiltonian reproduces the low-lying spectra and decay properties of fluorine and neon isotopes beyond the proton drip line.
We quantify many-body configuration and average partial-wave occupancies to elucidate the structural evolution of the drip line nuclei $^{12\text{--}14}$O, $^{13\text{--}15}$F, and $^{14\text{--}16}$Ne.
In particular, multi-proton separation energies and spectroscopic factors are analyzed in detail, leading to a prediction for the unresolved ground state of $^{13}$F. 
Furthermore, the candidate $4p$ emitter $^{14}$Ne is theoretically predicted for the first time, providing valuable guidance for future experimental investigations.

\end{abstract}


\end{frontmatter}

\section{Introduction}
Drip lines mark the stability limit against particle emission~\cite{SORLIN2008602,Ye2025,PhysRevLett.133.222501,PhysRevLett.108.142503,Ma2025}. Extremely proton-rich isotopes lying well beyond the proton drip line, such as $^{13}$F, $^{15}$Ne, and $^{16}$Ne, are unbound and decay through multi-proton emission, since no intermediate proton-bound isotopes exist along their proton-emission chains. These proton-emission processes provide valuable insight into the exotic nuclear structure~\cite{PhysRevLett.127.262502,hkmy-yfdk,WANG2026140030,Zhou2024}.

The lightest particle-stable nuclei of the oxygen, fluorine, and neon chains are $^{13}$O, $^{17}$F, and $^{17}$Ne, respectively.  Numerous isotopes lighter than those isotopes have been established in experiment.
In the fluorine chain, $^{15}$F, located two neutrons beyond the proton drip line, was first investigated in Refs~\cite{PhysRevC.17.1929,PhysRevC.17.1939}. 
Then, numerous experimental investigations of $^{15}$F were performed, including transfer reaction~\cite{LEPINESZILY2004331} and elastic resonance scattering in inverse kinematics~\cite{PhysRevC.68.034607,PhysRevC.69.031302,PhysRevC.72.034312,DEGRANCEY201626,PhysRevC.105.L051301}.
Further beyond the proton drip line, $^{14}$F was first explored through $^{13}\text{O}+p$ resonance scattering~\cite{GOLDBERG2010307}, and its structure was later investigated using the invariant-mass method~\cite{PhysRevC.107.054301}.
Experimentally, the $3p$ decay of $^{13}$F, which lies four neutrons beyond the proton drip line and ultimately leads to the particle-stable nucleus $^{10}$C, has been observed via a charge-exchange reaction ~\cite{PhysRevLett.126.132501}.
In the neon isotopic chain, $^{16}$Ne have been extensively studied through various reaction channels, such as neutron-knockout reactions from $^{17}$Ne ~\cite{PhysRevLett.113.232501}, the $^{20}\text{Ne}(^4\text{He},^8\text{Li})^{16}$Ne transfer reaction~\cite{PhysRevC.27.27,PhysRevC.17.1929}, and pion double-charge exchange reactions~\cite{PhysRevC.22.1180,PhysRevLett.79.3849}. 
The $^{15}$Ne was first observed by two-neutron knockout reactions from a $^{17}$Ne beam \cite{PhysRevLett.112.132502}.

Theoretical models have been employed to investigate unbound proton-rich F and Ne isotopes. 
Using the potential model, the low-lying resonances in $^{13}$F~\cite{PhysRevC.86.034301}, $^{14}$F~\cite{SHERR2011281} and $^{15}$F~\cite{PhysRevLett.99.089201} have been calculated.
The low-lying states of $^{15}$F have been extensively analyzed through elastic scattering reaction, using approaches such as the resonating group method~\cite{PhysRevC.72.024309}, the multichannel algebraic scattering theory~\cite{PhysRevLett.96.072502}, and the coupled-channel Gamow shell model~\cite{PhysRevC.105.L051301}. Moreover, the Gamow shell model (GSM) has also been applied to study $^{15}$F and $^{16}$Ne~\cite{PhysRevC.103.044319,sym17020169,PhysRevC.111.014302}.
Additional theoretical efforts include the $ab~initio$ no-core shell model calculation for $^{14}$F~\cite{PhysRevC.81.021301}, the $S$-matrix pole method for $^{15}$F ~\cite{PhysRevC.81.054314}, the analytic continuation in the coupling constant~\cite{ckxc-r2c3} for $^{16}$Ne, and the microscopic three-cluster model in the generator coordinate method for $^{15}$Ne~\cite{PhysRevC.99.064308}.
However, systematic investigations of remote isotopes beyond the drip line remain scarce.
To date, $^{14}$Ne has not been observed experimentally.
Furthermore, theoretical studies of nuclei in the vicinity of $^{13}$F remain scarce; in particular, microscopic calculations for $^{13}$F itself are virtually nonexistent. Therefore, applying state-of-the-art theoretical models to these extremely proton-rich systems is essential for understanding their exotic properties.

Nuclei beyond the drip lines belong to open quantum systems, for which particle emission channels are energetically accessible, and continuum coupling becomes unavoidable~\cite{Xu2024,PhysRevLett.127.262502,Michel_2009}. In such systems, continuum effects and inter-nucleon correlations are strongly intertwined and can qualitatively influence nuclear observables. Accurately incorporating these ingredients into a unified many-body description constitutes a major challenge for traditional bound-state nuclear structure models~\cite{Zhou2022,Jin2025}.

Among available theoretical approaches, the GSM~\cite{MichelSpringer,PhysRevLett.89.042502,PhysRevLett.97.110603,PhysRevC.84.051304,PhysRevC.96.054316,physics3040062} provides a powerful extension of the traditional shell model to the open quantum system. Building upon the Berggren ensemble~\cite{BERGGREN1968265,PhysRevC.47.768}, the GSM employs a completeness relation in the complex-energy plane. Both many-body correlations and continuum coupling are well treated in the GSM calculations.
Consequently, it is particularly well suited for weakly bound and unbound nuclei, where phenomena such as halo structures and resonances are prominent. Over the past few years, the GSM has been successfully applied to a wide range of unbound systems and drip line nuclei, demonstrating its predictive power for spectra, widths, and correlation patterns~\cite{Li2024,PhysRevC.106.L011301,PhysRevC.96.024308,PhysRevLett.89.042502,PhysRevC.67.054311,PhysRevC.111.034327,PhysRevC.96.024308,Li2024,8mt5-2k1z}.

In the present work, we perform systematic theoretical calculations for the proton-rich Ne, F, and O isotopes using the GSM.
The low-lying spectra and decay properties of $^{12-14}$O, $^{13-15}$F, and $^{15,16}$Ne are presented, along with the calculated multi-proton separation energies.
Moreover, we analyze the average occupations of different partial waves and the corresponding spectroscopic factors, revealing the role of continuum coupling in shaping the underlying configurations. 
Furthermore, our calculations predict the unbound nature of $^{14}$Ne, providing useful insights for future experimental investigations.

\section{Method}
The GSM employs the one-body Berggren basis~\cite{BERGGREN1968265,PhysRevC.47.768}, constructed from a finite-depth potential and comprising bound, resonant, and scattering states, which satisfies the completeness relation for each $(l, j)$ partial wave:
\begin{equation}\label{eq1}
\sum_{n}\left|u_{nlj}\right\rangle\left\langle u_{nlj}\right|+\int_{{L^{+}}}\left|u_{klj}\right\rangle\left\langle u_{klj}\right|dk=\mathbf{\hat{1}},
\end{equation}
where $\left|u_{nlj}\right\rangle$ represents a bound or resonant state, while $\left|u_{klj}\right\rangle$ denotes a scattering state along the complex contour $L^+$. All the discrete resonant states in Eq.~(\ref{eq1}) must be enclosed within the $L^+$ contour. 
To obtain an eigenproblem analogous to that of the traditional shell model, the $L^+$ contour is discretized using a Gauss–Legendre quadrature~\cite{Michel_2009}. 
The many-body completeness relation then follows by constructing Slater determinants from the one-body Berggren basis states. 
The GSM Hamiltonian matrix, which is complex symmetric, is diagonalized using the Jacobi-Davidson method~\cite{doi:10.1137/S0895479894270427,Sleijpen1996}. And the resonant states among the many-body continuum state are selected using the overlap approach~\cite{MICHEL2020106978,PhysRevC.67.054311}.
More details about the GSM framework have been thoroughly explained in Ref.~\cite{Michel_2009,MichelSpringer}.

The GSM works in a core + valence nucleons framework. The GSM Hamiltonian in the intrinsic nucleon-core cluster orbital shell model (COSM) coordinates~\cite{PhysRevC.38.410}, can be written as:
\begin{equation}\label{eq2}
\hat{H} = \sum_{i=1}^{N_{\text{val}}} \left(\frac{\hat{\mathbf{p}}_i^2}{2\mu_i} + \hat{U}_{\text{core}}(i) \right) + \sum_{i<j}^{N_{\text{val}}} \left( \hat{V}(i,j) + \frac{\hat{\mathbf{p}}_i \cdot \hat{\mathbf{p}}_j}{M_{\text{core}}} \right),
\end{equation}
where $N_{\text{val}}$ is the valence nucleons number, ${\mu_i}$ is the reduced mass of the $i$-th nucleon and $M_{\text{core}}$ is the mass of the core. The potential $\hat{U}_{\text{core}}$ denotes the one-body core-nucleon potential, and $\hat{V}(i,j)$ is the two-body interaction between valence nucleons. The last term represents the two-body recoil term. 

In this work, $^8$C is used as the inert core. While $^8$C is unbound, its decay width is very narrow (only 130 keV). In the studied nuclei ($^{12-14}$O, $^{13-15}$F, and $^{14-16}$Ne), at least two valence protons and two valence neutrons are present. The core's instability is then mitigated when valence nucleons are added.  So the $^8$C core is effectively bound and inert within a core + valence nucleons framework. The choice of an unbound core is not unprecedented in GSM, for example, $^{28}$O was used as the core in previous calculations of $^{39}$Mg~\cite{PhysRevC.94.054302}.
For proton-rich nuclei beyond $^8$C, it is sufficient to restrict the neutron valence space to the harmonic oscillator (HO) $p$ partial waves.
The proton valence space is represented by the Berggren basis, which contains the $S$-matrix poles $0p_{1/2}$, $1s_{1/2}$, and $0d_{5/2}$, as well as the complex contours of the $p_{1/2}$, $p_{3/2}$, $s_{1/2}$, and $d_{5/2}$ partial waves.
The $d_{3/2}$ partial wave is generated from the HO basis, as its larger centrifugal barrier leads to weaker continuum effects. 
The full configuration space is prohibitively large and needs to be truncated to render GSM calculations feasible. 
The GSM calculation using a natural orbital basis is adopted to reduce computational cost and improve convergence.
We first performed a GSM calculation within the Berggren basis, and the obtained one-body density matrix is diagonalized to generate the natural orbital basis~\cite{Brillouin1933}.
Then, we perform the GSM calculations within the constructed natural orbital basis. For calculations of $^{14-15}$Ne and $^{12-14}$O, at most four scattering states are allowed in the continuum (the 4$p$–4$h$ truncation) within the natural orbital basis, whereas $^{16}$Ne and $^{15}$F are carried out with a 3$p$–3$h$ truncation due to their higher computational demand. For $^{13-14}$F, the broad resonances are strongly embedded in the scattering continuum, for which the natural orbitals lead to unstable resonance identification, especially for the widths~\cite{PhysRevLett.119.032501}. We therefore adopt the Berggren basis with a 3$p$–3$h$ truncation.

To estimate the truncation uncertainty, we compared the results obtained in the $3p$--$3h$ and $4p$--$4h$ truncation spaces. The energy differences are only several tens of keV for $^{12-14}$O and remain below 300 keV for $^{14-16}$Ne. For the F isotopes, allowing one additional nucleon to occupy a scattering state changes the energy by only 12 keV, while allowing two additional nucleons in scattering states gives an energy difference below 120 keV. And the width uncertainty of F isotopes is below 100 keV.
The treatment properly incorporates continuum coupling and ensures reliable numerical accuracy.


\begin{table}[]
    \centering
    \setlength{\tabcolsep}{2.0mm}
    \renewcommand{\arraystretch}{1.1}     
    \caption{Binding energies (relative to the $^{8}$C inert core, in MeV) and widths (in keV) of the selected states of $A=11-15$ nuclei used to optimize the GSM Hamiltonian, with experimental values from~\cite{ensdf}.}
    \label{optimize}
    \begin{tabular}{cccccc}
    \hline\hline
    Nucleus & State & $E$ & $E_{\text{exp}}$ & $\Gamma$ & $\Gamma_{\text{exp}}$ \\
    \hline
    $^{11}$N & $1/2^+$ & $-$33.886 & $-$34.122 & 1343 & 830(30) \\
    $^{11}$N & $1/2^-$ & $-$33.400 & $-$33.392 & 459 & 600(100) \\
    $^{12}$N & $1^+$ & $-$49.046 & $-$49.224 &  &  \\
    $^{12}$O & $0^+$ & $-$33.701 & $-$33.768 &  &  \\
    $^{13}$O & $3/2^-$ & $-$50.753 & $-$50.727 &  &  \\
    $^{14}$O & $0^+$ & $-$74.094 & $-$73.912 &  &  \\
    $^{14}$F & $2^-$ & $-$49.087 & $-$49.174 & 617 & 910(100) \\
    $^{15}$F & $1/2^+$ & $-$72.561 & $-$72.659 & 265 & 660(20) \\
    $^{15}$Ne & $3/2^-$ & $-$48.563 & $-$48.280 &  & 590 \\
    \hline\hline
    \end{tabular}
\end{table}

\begin{table}[]
    \centering
    \setlength{\tabcolsep}{2.0mm}
    \renewcommand{\arraystretch}{1.1}     
    \caption{Optimized parameters of the EFT interaction at leading order (LO) and next-to-leading order (NLO) in natural units.}
    \label{parameter}
    \renewcommand{\arraystretch}{1.3} 
    \setlength{\tabcolsep}{2.6pt}       
    \begin{tabular}{lccccccc}
    \hline\hline
    LO constant & $C_{S0}$ & $C_{S1}$ &  &  &  &  & \\
    \hline
    LO value & $-$0.40 & $-$0.49 &  &  &  & & \\
    NLO constant & $C_1$ & $C_2$ & $C_3$ & $C_4$ & $C_5$ & $C_6$ & $C_7$ \\
    \hline
    NLO value & 0.43 & 4.42 & 0.78 & $-$3.64 & $-$4.26 & 1.81 & 3.66 \\
    \hline\hline
    \end{tabular}
\end{table}

The core potential is mimicked by a Woods-Saxon (WS) potential with a diffuseness of $d=0.65$ fm and a radius of $R_0=3$ fm.
The WS central potential depth $V_o$ is 42.5 MeV ($l=0$), 29.8 MeV ($l=1$), 38.8 MeV ($l=2$) for proton, and 51.3 MeV ($l=0,1$) for neutron.
The spin-orbit potential depth $V_{so}$ is fixed at 8.5 MeV for all $l>0$ partial waves.
The nucleon-nucleon interaction is described by the pionless effective field theory (EFT) interaction~\cite{RevModPhys.81.1773,RevModPhys.92.025004,RevModPhys.85.197,CONTESSI2017839,PhysRevC.92.054002}, including the leading-order (LO) and next-to-leading order (NLO) contact terms.
In this work, the interaction parameters are optimized by fitting the calculated binding energies of the low-lying states of selected nitrogen, oxygen, fluorine, and neon isotopes (see Table~\ref{optimize}) to the corresponding experimental values. 
The optimized EFT interaction parameters are listed in Table~\ref{parameter}. 
Note that within the pionless EFT framework, the LO interaction is characterized by the spin-independent ($C_S$) and spin-dependent ($C_T$) terms, which can be equivalently expressed as $C_{S0}=(C_S - 3C_T)$ and $C_{S1}=(C_S + C_T)$~\cite{PhysRevC.103.034305}.
In this work, we employ the representation in terms of $(C_{S0}, C_{S1})$ to describe the LO terms, which correspond to the $^1S_0$ and $^3S_1$ channels, respectively.
An $A$-dependence (0.25) on the number of nucleons has been added to the EFT two-body interaction to account for the effect of three-body interactions. The Coulomb interaction for protons is considered exactly, with charge radius $R_{\text{ch}}=3$ fm.
The root mean square deviation (RMSD) between the theoretical energies provided by the optimized GSM Hamiltonian and the experimental energies is 0.157 MeV.

\section{Results}

\begin{figure}[!htb]
\includegraphics[width=0.48\textwidth]{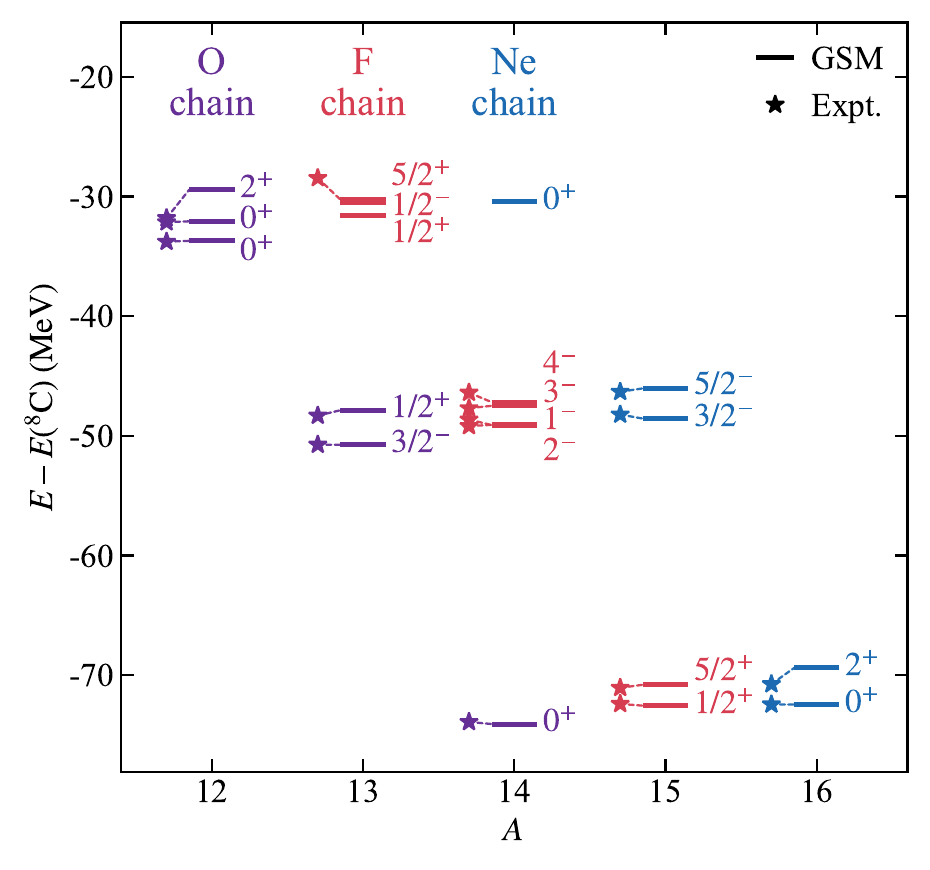}
\caption{\label{chain}
Spectra of oxygen, fluorine, and neon isotopic chains with $A=12-16$. 
All energies are given with respect to the $^8$C core. The energies are given in MeV.
Experimental data are taken from Ref.~\cite{PhysRevLett.126.132501} for $^{13}$F, Ref.~\cite{GOLDBERG2010307} for $^{14}$F, Ref.~\cite{PhysRevC.68.034607} for $^{15}$F, Ref.~\cite{PhysRevLett.112.132502} for $^{15}$Ne, and Refs.~\cite{PhysRevLett.113.232501,PhysRevC.92.034329} for $^{16}$Ne.
The low-lying states of the oxygen isotopes are taken from~\cite{ensdf}.}
\end{figure}

Based on the optimized GSM Hamiltonian, we have performed systematic GSM calculations for the proton-rich O, F, and Ne isotopes with $A=12-16$. The results are presented in Fig.~\ref{chain}. The GSM calculations reproduce the experimentally established low-lying states in the oxygen, fluorine, and neon isotopic chains with good accuracy.
The RMSD between the GSM energy and the corresponding experimental energy for the 17 reference states shown in the figure is 0.759 MeV, where many of them were not part of the optimization.
The largest discrepancy is 1.38~MeV for the $2^+$ state in $^{12}$O.

The GSM calculations are performed in a valence space including both pole states and scattering states, thereby including continuum coupling. To evaluate the effect of continuum coupling, we compare the GSM results with those obtained in the pole approximation, where only pole states are retained. Continuum coupling reduces the ground-state separation energies by 0.418 MeV in $^{15}$F, 0.704 MeV in $^{14}$F, and 0.814 MeV in $^{13}$F, demonstrating its sizable impact on these proton-rich systems.

Furthermore, we have calculated the proton and neutron partial-wave occupancies for the low-lying states of $^{12-14}$O, $^{13-15}$F, and $^{14-16}$Ne, as presented in Table~\ref{Occupation-tab}.
The average occupations reveal a systematic structural evolution. Along the $N=4$, 5, and 6 isotonic chains, the proton $1s_{1/2}$ occupancy increases from the O isotopes to the F and Ne isotopes. Within each nucleus, most higher-lying excited states exhibit enhanced $0d_{5/2}$ occupancy relative to the ground state or the lowest-lying states, as seen in $^{15}$F$(5/2^+)$, $^{16}$Ne$(2^+)$, $^{14}$F$(3^{-}, 4^{-})$, $^{15}$Ne$(5/2^-)$, and $^{13}$F$(5/2^+)$. These trends point to a growing competition between the $1s_{1/2}$ and $0d_{5/2}$ configurations of extremely proton-rich nuclei, indicating stronger orbital rearrangement and $sd$-shell mixing beyond the proton drip line. 

States with dominant proton $1s_{1/2}$ components couple more strongly to the continuum, since the $s$-wave carries $l=0$ and experiences no centrifugal barrier. They therefore tend to appear as broader resonances. By contrast, states with dominant $0d_{5/2}$ character are hindered by the $l=2$ centrifugal barrier and remain comparatively narrower. The widths are clearly demonstrated in the detailed analysis of each O-F-Ne isotonic chain presented below.

\begin{table}[]
    \centering
    \setlength{\tabcolsep}{1.0mm}
     \renewcommand{\arraystretch}{1.1}     
    \caption{The calculated average occupations of the proton and neutron partial waves for the low-lying states. }
    \label{Occupation-tab}
    \begin{tabular}{ccccccccc}
    \hline\hline
    Nucleus& State & $\pi p_{3/2}$ & $\pi p_{1/2}$ & $\pi s_{1/2}$ & $\pi d_{5/2}$ & $\pi d_{3/2}$ &$\nu p_{3/2}$ & $\nu p_{1/2}$\\
    \hline
    $^{14}$O & $0_1^+$ & 0.01 & 1.44 & 0.24 & 0.27 & 0.04 & 3.83 & 0.17\\
    $^{15}$F & $1/2_1^+$ & 0.01 & 1.59 & 1.10 & 0.27 & 0.04 & 3.84 & 0.16\\
     & $5/2^+_1$ & 0.01 & 1.61 & 0.18 & 1.16 & 0.04 & 3.82 & 0.18\\
    $^{16}$Ne & $0^+_1$ & 0.02 & 1.66 & 1.59 & 0.67 & 0.06 & 3.84 & 0.16\\
     & $2^+_1$ & 0.01 & 1.78 & 0.97 & 1.19 & 0.04 & 3.83 & 0.17\\   
    \hline
    $^{13}$O & $3/2_1^-$ & 0.02 & 1.32 & 0.30 & 0.31 & 0.05 & 2.91 & 0.09\\
     & $1/2^+_1$ & 0.01 & 0.99 & 0.97 & 0.03 & 0.00 & 2.91 & 0.09\\
    $^{14}$F & $2_1^-$ & 0.02 & 1.58 & 1.02 & 0.32 & 0.06 & 2.92 & 0.08\\
      & $1_1^-$ & 0.02 & 1.60 & 0.99 & 0.34 & 0.06 & 2.92 & 0.08\\
      & $3_1^-$ & 0.02 & 1.50 & 0.27 & 1.17 & 0.04 & 2.91 & 0.09\\
      & $4_1^-$ & 0.01 & 1.45 & 0.28 & 1.21 & 0.04 & 2.91 & 0.09\\
    $^{15}$Ne & $3/2^-_1$ & 0.03 & 1.59 & 1.60 & 0.70 & 0.08& 2.92 & 0.08\\
     & $5/2^-_1$ & 0.02 & 1.74 & 0.96 & 1.22 & 0.06 & 2.91 & 0.09\\
    \hline
     $^{12}$O & $0_1^+$ & 0.02 & 1.13 & 0.45 & 0.35 & 0.05 & 1.87 & 0.13\\
     & $0_2^+$ & 0.004 & 0.61 & 1.37 & 0.02 & 0.002 & 1.87 & 0.13\\
    $^{13}$F & $1/2_1^+$ & 0.01 & 1.42 & 1.17 & 0.34 & 0.05 & 1.88 & 0.12\\
     & $1/2_1^-$ & 0.01 & 1.62 & 1.13 & 0.21 & 0.02 & 1.88 & 0.12\\
     & $5/2_1^+$ & 0.01 & 1.15 & 0.51 & 1.29 & 0.03 & 1.88 & 0.12\\
    $^{14}$Ne & $0^+_1$ & 0.03 & 1.55 & 1.68 & 0.66 & 0.08 & 1.88 & 0.12\\
    \hline\hline
    \end{tabular}
\end{table}

\begin{figure}[!htb]
\includegraphics[width=0.42\textwidth]{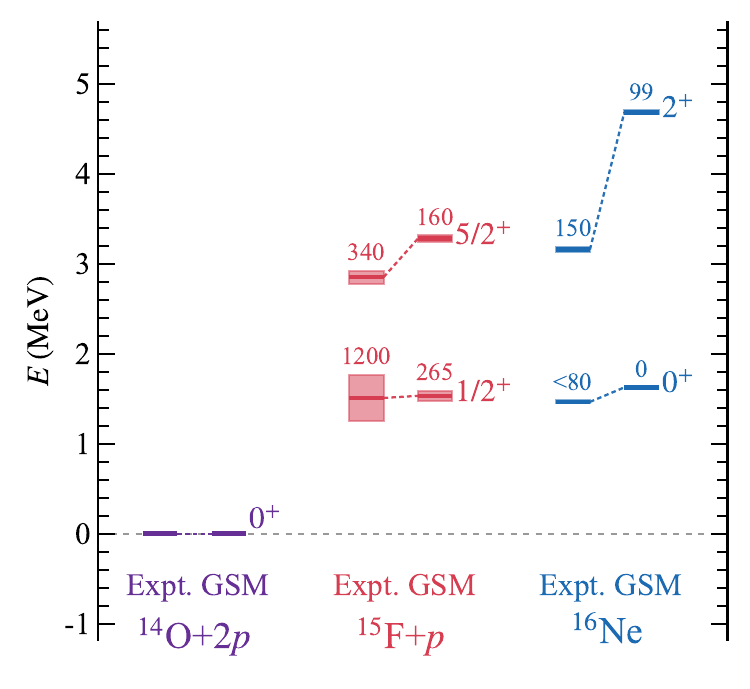}
\caption{\label{15F}
Calculated low-lying states in $^{16}$Ne, $^{15}$F and $^{14}$O, compared with experimental data taken from Ref.~\cite{PhysRevC.68.034607} for $^{15}$F and Refs.~\cite{PhysRevLett.113.232501,PhysRevC.92.034329} for $^{16}$Ne.
Energies are given in MeV relative to the ground state of $^{14}$O.
For the resonant state, the box width indicates the decay width ($\Gamma$) in units of keV.}
\end{figure}

We next examine each proton-emission chain in detail to explore their distinct structural properties.
Figure~\ref{15F} shows the calculated level schemes of the $N=6$ isotones, with energies referenced to the $^{14}$O+$2p$ and $^{14}$O+$p$ thresholds for $^{16}$Ne and $^{15}$F, respectively.

The low-lying structure of $^{15}$F is now well studied~\cite{PhysRevC.72.034312,PhysRevC.69.031302,PhysRevC.68.034607,LEPINESZILY2004331,PhysRevC.17.1929,PhysRevC.17.1939}: the ground state, $J^{\pi}=1/2^+$, lies about 1.2-1.6~MeV above the $^{14}$O$+p$ threshold and appears as a broad resonance ($\Gamma\approx0.5$–$1.3$~MeV), whereas the first excited state $5/2^+$ is unbound by about 2.8~MeV and is comparatively narrow ($\Gamma\approx0.3$~MeV).
In our GSM calculation, the $1/2^+$ and $5/2^+$ states of $^{15}$F are about 1.53 and 3.28 MeV unbound by $1p$ decay, respectively. 
Our calculations reproduce the separation energy well. However, relative to existing data, the excitation energy is slightly overestimated by about 500 keV, and the decay widths of the two states are underestimated.
For $^{15}$F, the ground state is interpreted as mainly $\pi\{(1s_{1/2})^1(0p_{1/2})^2\}\otimes\nu(0p_{3/2})^4$, and the first excited state as $\pi\{(0p_{1/2})^2(0d_{5/2})^1\}\otimes\nu(0p_{3/2})^4$.
The structure of the ground state can be interpreted as a proton orbiting with $l = 0$ above a $^{14}$O$_{g.s.}$ core.
Moreover, GSM calculation yields a spectroscopic factor $C^2S (s_{1/2})=0.92$ for the $^{15}$F ground state ($1/2^{+}$), in close agreement with the IMME-based estimate $C^2S=0.88$ by H. T. Fortune~\cite{PhysRevC.74.054310}. For the $5/2^{+}$ state we obtain $C^2S (d_{5/2})=0.94$, consistent with $C^2S=0.92(12)$ extracted from the transfer reaction on neon~\cite{PhysRevC.17.1939}. These values corroborate the good single-particle character of the $1/2^{+}$ and $5/2^{+}$ states in $^{15}$F.
To further examine mirror symmetry breaking and continuum effects, we compare the low-lying spectra of $^{15}$F and its mirror partner $^{15}$C. The GSM calculation reproduces the mirror spectrum, although the calculated $5/2^+$ excitation energy is about 0.48 MeV higher than the experimental value in $^{15}$C. The mirror energy difference of the $5/2^+$ state obtained in our GSM is 526 keV, which is close to the experimental values of 603 keV.

For $^{16}$Ne, experimental studies indicate that the ground state lies about 1.34-1.47~MeV above the$^{14}$O$_{g.s.}+2p$ threshold, with uncertainties of several tens of keV, and first excited state lies at about 3.16 MeV above the same threshold with a narrow width~\cite{PhysRevC.17.1929,PhysRevC.27.27,PhysRevC.22.1180,PhysRevC.82.054315,PhysRevLett.112.132502,PhysRevLett.113.232501}. 
For $^{16}$Ne, the ground state is unbound by 1.627~MeV with respect to the $^{14}$O$+2p$ threshold in GSM calculation, close to the experimental value of 1.466(20)~MeV~\cite{PhysRevLett.113.232501}.
The first $2^{+}$ state is predicted at 4.685~MeV, whereas experiment reports 3.16(2)~MeV~\cite{PhysRevC.92.034329}, with a deviation of $\approx$1.5~MeV.
Structurally, the $^{16}\text{Ne}_{g.s.}$ is dominated by $\pi\{(1s_{1/2})^2(0p_{1/2})^2\}\otimes\nu(0p_{3/2})^4$, with small admixtures of $\pi\{(0p_{1/2})^2(0d_{5/2})^2\}\otimes\nu(0p_{3/2})^4$ (10\%) and $\pi\{(1s_{1/2})^2(0d_{5/2})^2\}\otimes\nu(0p_{3/2})^4$ (7\%). While the first $2^+$ state is primarily $\pi\{(0p_{1/2})^2(1s_{1/2})^1(0d_{5/2})^1\}\otimes\nu(0p_{3/2})^4$.

\begin{figure}[!htb]
\includegraphics[width=0.42\textwidth]{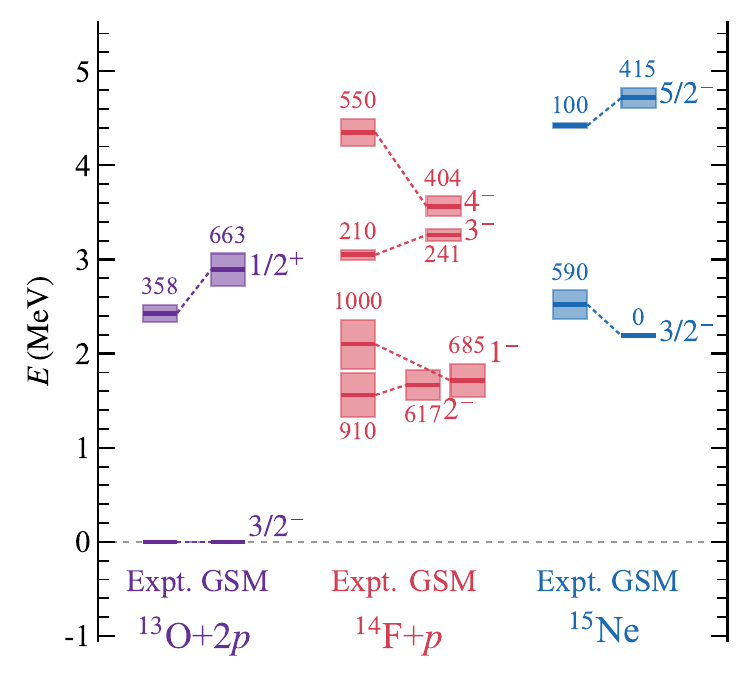}
\caption{\label{14F}
The spectra of low-lying states in $^{15}$Ne, $^{14}$F and $^{13}$O, shown relative to the ground state of $^{13}$O. The presentation is similar to that of Fig.~\ref{15F}. Experimental data are taken from Ref.~\cite{GOLDBERG2010307} for $^{14}$F and Ref.~\cite{PhysRevLett.112.132502} for $^{15}$Ne.}
\end{figure}

Having established the $N=6$ benchmarks ($^{14}$O–$^{15}$F–$^{16}$Ne), we now turn to the more neutron-deficient $N=5$ chain ($^{13}$O–$^{14}$F–$^{15}$Ne).
As shown in Fig.~\ref{14F}, the energies are plotted relative to the ground state of $^{13}$O.
$^{13}$O is the lightest particle-stable oxygen isotope, whereas $^{14}$F and $^{15}$Ne are unbound and will decay to $^{13}$O via $1p$ and $2p$ emission, respectively.

For $^{14}$F, our GSM calculation gives the single-proton separation energy $S_p=1.66$ MeV, highly consistent with the experimental value of 1.56 MeV~\cite{GOLDBERG2010307}.
We place the $3^{-}$ level about 0.2~MeV above the assignment of Ref.~\cite{GOLDBERG2010307}, close to an experimentally unobserved region; consistently, the invariant-mass analysis locates this resonance $\sim$0.5~MeV higher~\cite{PhysRevC.107.054301}, and the estimate of Sherr and Fortune is $\sim$0.6~MeV higher~\cite{SHERR2011281}. 
The calculated energies for the $1^-$ and $4^-$ states are slightly underestimated relative to the experiment, with a maximum deviation of 0.8~MeV.
Our calculation gives the main configuration of $^{13}$O$_{g.s.}$ as: $\pi(0p_{1/2})^2\otimes\nu(0p_{3/2})^3$. For $^{14}$F, the first two states are dominant by $\pi\{(1s_{1/2})^1(0p_{1/2})^2\}\otimes\nu(0p_{3/2})^3$. 
For the narrower states $4^-$ and $3^-$, the main configurations are $\pi\{(0p_{1/2})^2(0d_{5/2})^1\}\otimes\nu(0p_{3/2})^3$. As the excitation energy increases, valence protons in the $s_{1/2}$ orbital tend to be excited to the $d_{5/2}$ orbital.

We now compare the spectroscopic factors from our GSM calculation with those of Sherr and Fortune in Ref.~\cite{SHERR2011281}:  
For the ground state $2^{-}_{\mathrm{g.s.}}$, both calculations indicate a nearly pure $s$-wave resonance: $C^2S=0.88$ in this work versus $\sim0.98$ of Sherr and Fortune.  
For the $1^{-}$ state, we obtain $C^2S=0.85$, notably larger than predicted $\sim0.55$, implying a stronger $s$–wave single–particle strength in our model, yet still less pure than $2^{-}_{\mathrm{g.s.}}$.
For the $3^{-}$ and $4^{-}$, the spectroscopic factors to the $^{13}$O$_{\mathrm{g.s.}}+d_{5/2}$ channel are 0.93 and 0.96, respectively—well above the predicted $0.62(12)$ and $0.54(10)$.
Our calculation supports a cleaner $d_{5/2}$ single-particle character. 
These discrepancies likely reflect differences in the treatment of continuum coupling and the choice of potential parameters, indicating the model sensitivity of extracted spectroscopic strengths.

\begin{figure}[!htb]
\centering
\includegraphics[width=0.34\textwidth]{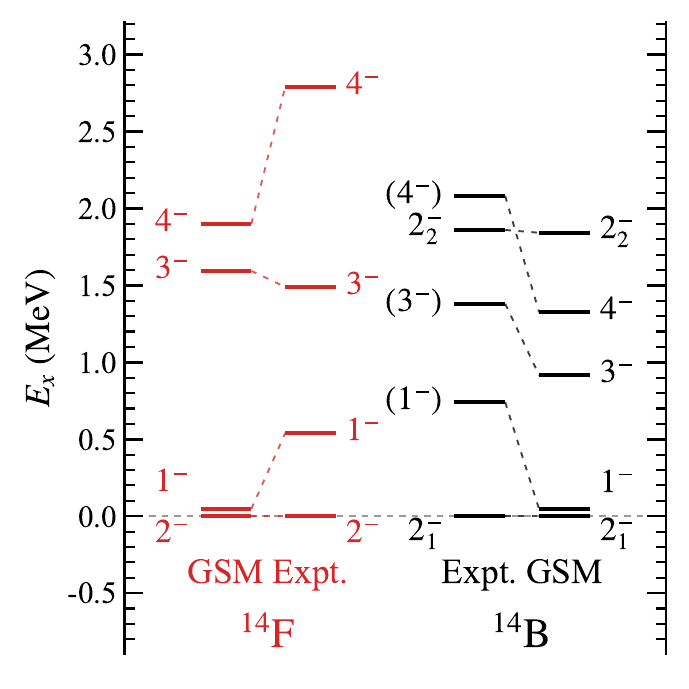}
\caption{\label{14B}The spectra of low-lying states in mirror pair $^{14}$F/$^{14}$B. The excitation energies are related to the ground state. Experimental data of $^{14}$B is taken from Ref.~\cite{AJZENBERGSELOVE19911}.}
\end{figure}

The mirror-symmetry breaking in the $^{14}$F/$^{14}$B mirror pair is also investigated, as shown in Fig.~\ref{14B}. Overall, the GSM calculation reproduces the low-lying spectra reasonably well. The calculated orbital occupations and dominant configurations of the mirror states are nearly identical, indicating that these states have very similar underlying structures. The remaining differences mainly arise from continuum coupling effects associated with the extended $s$-wave component, which are more pronounced in the proton-rich nucleus $^{14}$F. As a result, the calculated excitation energies of the $3^-$ and $4^-$ in $^{14}$F are shifted higher than their counterparts in $^{14}$B. The results show that continuum coupling provides a non-negligible binding contribution on the $s$-wave component in the proton-rich unbound nucleus, consistent with the situation in the mirror pair $^{16}$F/$^{16}$N~\cite{PhysRevC.106.L011301}.

The $^{15}$Ne was studied in two-neutron knockout reactions, and the ground state is unbound by 2.522(66) MeV above the $^{13}\text{O} + 2p$ decay threshold~\cite{PhysRevLett.112.132502}.
Our GSM calculation predicts the $^{15}$Ne ground state to be unbound by 2.19~MeV with respect to the $^{13}$O$+2p$ threshold and places the $5/2^{-}$ level 4.72~MeV above $^{13}$O$_{\mathrm{g.s.}}$. 
The calculated $^{15}$Ne spectrum agrees with experimental data within 400~keV. 
The GSM calculation gives that the ground state $^{15}$Ne is dominated by the proton configuration $\pi\{(1s_{1/2})^{2}(0p_{1/2})^{2}\}\!\otimes\!\nu(0p_{3/2})^{3}$ (44.2\%), with smaller admixtures $\pi\{(1s_{1/2})^{2}(0d_{5/2})^{2}\}\!\otimes\!\nu(0p_{3/2})^{3}$ (10.3\%) and $\pi\{(0p_{1/2})^{2}(0d_{5/2})^{2}\}\!\otimes\!\nu(0p_{3/2})^{3}$ (10.3\%); the residual strength is among the higher scattering components. 
Average occupancies in Table~\ref{Occupation-tab} indicate a $^{13}$O core coupled mainly to $1s_{1/2}$ protons with a modest $0d_{5/2}$ admixture, consistent with the experimental $s$-wave fraction of 63(5)\%~\cite{PhysRevLett.112.132502}. 
For the $5/2^{-}$ excited state, strength is redistributed toward the $0d_{5/2}$ orbital with a slight reduction of the $1s_{1/2}$ component.

\begin{figure}[!htb]
\includegraphics[width=0.48\textwidth]{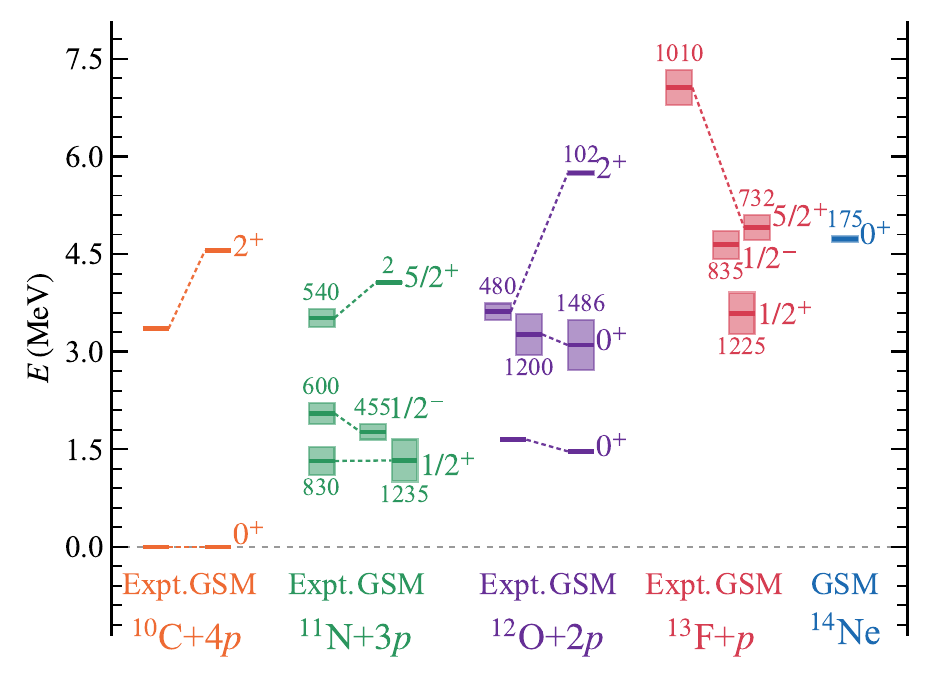}
\caption{\label{13F}
The spectra of low-lying states in $^{14}$Ne, $^{13}$F, $^{12}$O, $^{11}$N and $^{10}$C, shown relative to the $^{10}$C ground state. The presentation is similar to that of Fig.~\ref{15F}. Experimental data are taken from Ref.~\cite{PhysRevLett.126.132501}.}
\end{figure}

For the more proton-rich $N=4$ isotonic chain, $^{12}$O is unbound with respect to the $^{10}$C$+2p$ threshold and decays by $2p$ emission.
Figure~\ref{13F} presents the calculated and experimental low-lying spectra of the $N=4$ isotones, with energies referenced to the $^{10}\text{C}_\text{g.s.}$. Experimental data of $^{10}$C, $^{11}$N and $^{12}$O are taken from~\cite{ensdf}.

For $^{13}$F, a resonance was observed in the invariant-mass study and probably corresponds to a $5/2^+$ excited state~\cite{PhysRevLett.126.132501}.
In this work, our GSM calculations give three low-lying states of $^{13}$F, including the observed excited state $5/2^+$, the expected ground state $1/2^+$, and a candidate $1/2^{-}$ excitation.
GSM calculations predict the ground state $1/2^+$ is located 3.59 MeV above the threshold for $3p$ decay to $^{10}\text{C}_{\text{g.s.}}$, with decay width of 1.225 MeV. For the $1/2^-$ excited state, the decay energy is about 4.64 MeV with $\Gamma=0.835$ MeV. And for the $5/2^+$ excited state, the decay energy is about 4.91~MeV with $\Gamma=0.732$~MeV, about 2.15~MeV lower and markedly narrower than the experimental value~\cite{PhysRevLett.126.132501} [$E=7.06(9)$~MeV, $\Gamma=1.01(27)$~MeV].
Relative to the $^{12}$O$+p$ threshold, our GSM calculation yields $E_r(1/2^+)=2.12$~MeV, $E_r(1/2^-)=3.18$~MeV, and $E_r(5/2^+)=3.44$~MeV. 
The estimates of Fortune and Sherr~\cite{PhysRevC.86.034301} give $E_r(1/2^+)=2.40$~MeV with $\Gamma=0.60(11)$~MeV and $E_r(5/2^+)=4.94$ (or 5.26)~MeV with $\Gamma=0.3$--$0.4$~MeV.
Our results place the $1/2^+$ and $5/2^+$ states lower by $\sim$0.28~MeV and $\sim$1.5--1.8~MeV, respectively, compared to that results in Ref.~\cite{PhysRevC.86.034301}.
We obtain a $1/2^--5/2^+$ spacing of $\approx$0.27~MeV, and Ref.~\cite{PhysRevC.86.034301} anticipated the separation in $^{13}$F comparable to that in $^{13}$Be. 

In $^{13}$F, the $1/2^+$ state is primarily $\pi\{(1s_{1/2})^1(0p_{1/2})^2\}\otimes\nu(0p_{3/2})^2$ (35.6\%), with a admixture of $\pi\{(0p_{1/2})^2(\widetilde{s}_{1/2})^1\}\otimes\nu(0p_{3/2})^2$ (19.8\%).
The orbital occupation (Table~\ref{Occupation-tab}) likewise indicates the $1/2^+$ state has a configuration corresponding to a single valence proton in the $1s_{1/2}$ orbital outside the $^{12}\text{O}_{\text{g.s.}}$ core.
Accordingly, the GSM calculation for the spectroscopic factor of  $^{12}$O$_{\mathrm{g.s.}}+s_{1/2}$ provides a large value, as $C^2S= 0.89$.
The excited $1/2^-$ state is dominated by $\pi\{(1s_{1/2})^2(0p_{1/2})^1\}\otimes\nu(0p_{3/2})^2$.
For the $5/2^+$ excited state, the enhanced $0d_{5/2}$ occupation, as shown in Table~\ref{Occupation-tab}, contains sizable $d_{5/2}$ scattering components, indicating stronger coupling to the continuum.
The spectroscopic factor of the $1/2^-$ state for the $^{12}$O$_{\mathrm{g.s.}}+p_{1/2}$ channel is 0.84, while it is 0.93 for the $5/2^+$ state in the $^{12}$O$_{\mathrm{g.s.}}+d_{5/2}$ channel. 


Additionally, the GSM calculation allows one to assess the ground state of the unbound $^{14}$Ne, which has not yet been observed experimentally and is expected to be unbound with respect to $4p$ emission. The $^{14}\text{Ne}_{g.s.}$ is predicted to be 4.73 MeV above the $^{10}\text{C}+4p$ threshold, with a decay width of 175 keV. The main configuration is $\pi\{(1s_{1/2})^2(0p_{1/2})^2\}\otimes\nu(0p_{3/2})^2$ (50.0\%), with a small admixture of $\pi\{(1s_{1/2})^2(0d_{5/2})^2\}\otimes\nu(0p_{3/2})^2$ (13.2\%). 
Nevertheless, the spectroscopic factors suggest a possible structural origin of the predicted narrow width. The dominant one-proton overlap is associated with the hindered $^{13}$F$(1/2^-)+p_{1/2}$ channel, for which the decay energy is small because the $^{13}$F$(1/2^-)$ state lies close to the $^{14}$Ne ground state.

In GSM calculations, the calculated widths can be affected by both model-space truncation and the use of natural orbitals, especially for nuclei with several valence nucleons, such as $^{15}$Ne and $^{14}$Ne. In these cases, the natural orbitals are generated from a strongly truncated Berggren basis because of the rapidly increasing computational cost, which may affect the decay widths but the corresponding energies remain rather stable.
Thus, the narrow width predicted for the $^{14}$Ne ground state should be viewed as a model prediction rather than a high-precision result, with the qualitative implication that $^{14}$Ne is a promising candidate for $4p$ emission.

\section{Summary}
Systematic investigations of nuclei far beyond the proton drip line in the vicinity of $^{13}$F, where multi-proton emission is expected, remain scarce.
In the present work, we address these open quantum systems within GSM, in which the continuum coupling and inter-nucleon correlations are both well treated, to carry out a comprehensive theoretical study.

The GSM Hamiltonian, based on a pionless effective field theory nucleon–nucleon interaction, is optimized, yielding an RMSD of 0.157~MeV for the fitted levels. 
With this Hamiltonian, we perform a comprehensive study of the low-lying spectra, decay widths, and multi-proton separation energies of $^{12\text{--}14}$O, $^{13\text{--}15}$F, and $^{14\text{--}16}$Ne, focusing in particular on the $3p$ emitter $^{13}$F, the $2p$ emitter $^{15}$Ne, and the candidate $4p$ emitter $^{14}$Ne. 
Across the oxygen, fluorine, and neon chains with $A=12\text{--}16$, the calculations reproduce the experimental spectra with an overall RMSD of 0.759~MeV.
The proton separation energies are well reproduced for $^{15}$F and $^{14}$F: $S_p(^{15}\mathrm{F})=1.53$~MeV and $S_p(^{14}\mathrm{F})=1.67$~MeV, in accord with the experimental values of 1.51(11) and 1.56(4)~MeV, respectively.
For $^{13}$F, we confirm the observed $5/2^{+}$ resonance as an excited state with a decay energy of \textcolor{blue}{4.91}~MeV relative to the $^{10}$C$+3p$ threshold. 
We further predict a $1/2^{+}$ ground state of $^{13}$F at $\sim$\textcolor{blue}{3.59}~MeV above the $^{10}$C$+3p$ threshold and unbound by $\sim$\textcolor{blue}{2.12}~MeV with respect to $^{12}$O$+p$. 
The partial-wave occupancies and spectroscopic factors highlight proton-driven structural evolution and the essential role of continuum coupling in intruder states and configuration mixing.
Moreover, the ground state of the experimentally unobserved $^{14}$Ne is predicted to lie $\sim$4.73~MeV above the $^{10}$C$+4p$ threshold. Therefore, it is predicted to be a $4p$ emitter, providing useful insights for future experimental investigations.

\section*{Acknowledgements}
This work has been supported by the National Key R\&D Program of China under Grant No. 2023YFA1606403, 2024YFE0109800, and 2024YFE0109802; the National Natural Science Foundation of China under Grant Nos. 12205340, 12175281, 12347106, 12575124, 12441506, and 12405141; the Gansu Natural Science Foundation under Grant Nos. 22JR5RA123 and 25JRRA467. The numerical calculations in this paper have been done at Hefei Advanced Computing Center.

\section*{Reference}

\bibliographystyle{elsarticle-num_noURL}

\bibliography{Ref}

\end{document}